\documentclass[conference]{IEEEtran}
\IEEEoverridecommandlockouts
\usepackage[hyphens]{url}  
\usepackage{graphicx} 
\usepackage{natbib}  
\usepackage{caption} 
\usepackage{algorithm}
\usepackage{algorithmic}

\usepackage{multirow}
\usepackage{booktabs}
\usepackage{graphicx}
\usepackage{caption}
\usepackage{makecell}
\usepackage{amssymb}
\usepackage{amsmath}
\usepackage{marvosym}

\usepackage{tcolorbox}
\usepackage[table]{xcolor}
\tcbuselibrary{skins}

\newenvironment{mybox}{\begin{tcolorbox}}{\end{tcolorbox}}

\newcommand{{\sysname}}{\texttt{Moirae}}

\usepackage{newfloat}
\usepackage{listings}
\DeclareCaptionStyle{ruled}{labelfont=normalfont,labelsep=colon,strut=off} 
\floatstyle{ruled}
\newfloat{listing}{tb}{lst}{}
\floatname{listing}{Listing}

\usepackage{booktabs}

\title{{\sysname}: A Multimodal Agent Collaborative Framework for Dynamic Android Malware Detection}

\author{
Xueying Zeng$^\star$\textsuperscript{$\dagger$}\thanks{\textsuperscript{$\dagger$} Equal contribution.},
Youquan Xian$^\S$\textsuperscript{$\dagger$},
Yanze Li$^\star$,
Bowen Hu$^\star$,
Ziqi Shan$^\star$,
Xu Luo$^\star$,
Danping Yang$^\ddagger$, \\
Peng Liu$^\ddagger$,
Lei Cui$^\S$,
Bo Li$^{\star}$\textsuperscript{\Letter}\thanks{\textsuperscript{\Letter} Corresponding author.}
\\
\\
$^{\star}$ School of Computer Science and Engineering, Beihang University, Beijing, China \\
$^{\S}$ School of Cyberspace Security, Beijing University of Posts and Telecommunications, Beijing, China \\
$^{\ddagger}$ School of Computer Science and Engineering, Guangxi Normal University, Guilin, China
}

\begin{document}

\maketitle

\begin{abstract}
The Android ecosystem faces persistent and rapidly evolving malware threats. Existing machine learning detectors are vulnerable to concept drift because they rely on implementation-specific features whose distributions change over time. Large language models (LLMs) offer strong semantic understanding and zero-shot reasoning, but current LLM-based detectors typically depend on code-centric or single-dimensional evidence, making them susceptible to obfuscation and limiting comprehensive behavior analysis.
We present {\sysname}, a multimodal agent collaborative framework for dynamic Android malware detection. {\sysname} dynamically collects multimodal runtime evidence and employs ReAct-based specialized agents to analyze complementary behavioral views.
The detection process begins by identifying visual deception cues, modeling UI state transitions, and integrating runtime API behaviors to fuse multi-dimensional evidence across user-visible interfaces and hidden backend operations.
Experiments on temporally and distributionally unseen datasets show that {\sysname} achieves an accuracy of 90.06\% without fine-tuning, outperforming state-of-the-art baselines and demonstrating strong zero-shot generalization against Android malware concept drift.

\end{abstract}



\section{Introduction}
Android dominates the mobile ecosystem, making it a major target for malware attacks and posing a persistent security threat to hundreds of millions of users \cite{faruki2014android, bhat2019survey}. Most automated Android malware detectors rely on static or dynamic features and supervised models that learn statistical correlations between observable program artifacts and malware labels \cite{aafer2013droidapiminer, li2018significant, wu2019malscan}. These models typically assume that training and test data follow similar distributions. In practice, however, benign and malicious applications continuously evolve, causing their feature distributions to shift over time, particularly in shallow feature spaces \cite{tang2023demystifying}. Consequently, detectors trained on historical samples often exhibit substantial performance degradation on emerging applications, limiting their reliability in long-term deployment.

Prior work has sought to mitigate concept drift through Markov-chain modeling \cite{onwuzurike2019mamadroid}, semantic-embedding-based clustering \cite{xu2020sdac, yang2024novel}, and adaptive distribution tracking \cite{yang2021cade, fernando2024fesad}. Dynamic analysis has also been widely adopted to capture runtime behaviors and execution logic that may evade static inspection. Such methods collect system-call traces, network traffic, and runtime API sequences in controlled environments \cite{tang2024dtdroid, huang2024strengthening, feng2025hgdetector}. More recent approaches use pseudo-labeling \cite{xu2019droidevolver} or continuous active learning \cite{chen2023continuous} to update detection models with newly collected samples. Although these techniques can partially alleviate concept drift, they remain largely dependent on learning feature-label correlations from evolving data distributions. They therefore require continuous sample collection, annotation, and model retraining, while offering limited understanding of the high-level intent underlying observed behaviors.

Large language models (LLMs) offer a promising alternative due to their strong knowledge transfer, zero-shot generalization, and semantic reasoning capabilities \cite{islam2025gpt, glm2024chatglm}. Existing LLM-based Android malware detectors generally follow two paradigms: using LLMs to summarize program features for downstream classifiers \cite{zhao2025apppoet, yan2025prompt}, or directly reasoning over decompiled code snippets \cite{he2025benchmarking, qian2025lamd}. The former treats LLMs merely as feature extractors and overlooks their ability to reason about behavioral intent. The latter relies on large-scale code analysis, which is vulnerable to code obfuscation. More fundamentally, both paradigms typically rely on a single, code-centric source of evidence. Such evidence provides only a partial view of an application's behavior and may be insufficient to distinguish malicious actions from legitimate functionality.



Our key observation is that Android malicious behavior manifests across multiple complementary semantic levels. Visual content reveals how an application presents itself and induces user actions, UI interactions capture state transitions and user responses, and system-level API operations expose the resulting sensitive consequences. No single modality is sufficient: suspicious interfaces do not reveal what the application ultimately executes, while isolated API calls may appear benign without their user-facing context. For example, sending an SMS may be legitimate in a messaging application but suspicious when triggered after a deceptive prompt without clear user consent. By linking what users see, how they interact, and what the application performs, cross-level evidence provides a more complete account of malicious behavior. Although interfaces, code structures, and API combinations may evolve, the underlying malicious objectives, such as deceptive advertising, user intimidation, SMS abuse, and extortion, remain comparatively stable. This motivates multimodal reasoning to infer high-level malicious intent from evolving low-level implementations.

Motivated by this observation, we propose {\sysname}\footnote{{\sysname}, named after the three Greek goddesses who control the threads of fate, symbolizes our framework's ability to weave together diverse behavioral threads—visual content, UI interactions, and API operations—to connect low-level execution with high-level intent.}, an LLM-powered multimodal agent framework for Android malware detection. Through dynamic execution, {\sysname} collects runtime evidence from three complementary dimensions: visual inducements presented by the application, UI states and interaction transitions, and system-level API operations. It aligns these observations across semantic levels to connect user-visible behavior and interactions with their underlying system consequences. Multiple specialized LLM agents then analyze and integrate the resulting evidence to infer high-level malicious intent. By exploiting the multimodal understanding and reasoning capabilities of LLMs, {\sysname} moves beyond implementation-specific feature fitting and enables more comprehensive analysis of previously unseen applications without fine-tuning.
The main contributions of this paper are as follows:

\begin{itemize}
    \item We proposed {\sysname}. By dynamically capturing and fusing runtime evidence across three complementary dimensions, visual presentation, UI interaction transitions, and system-level API operations, {\sysname} effectively correlates user-visible behaviors with underlying system consequences.
    
    \item We designed a multi-agent autonomous analysis mechanism based on the ReAct paradigm. Specialized LLM agents collaboratively process the cross-level evidence, enabling the framework to autonomously align and mutually verify multimodal data.
    
    \item Experimental evaluations demonstrate that {\sysname} achieved a 90.06\% accuracy. Furthermore, compared to the SOTA baseline, the framework exhibits excellent generalization performance.
\end{itemize}

\begin{figure}[htbp]
\centering
\includegraphics[width=0.95\linewidth]{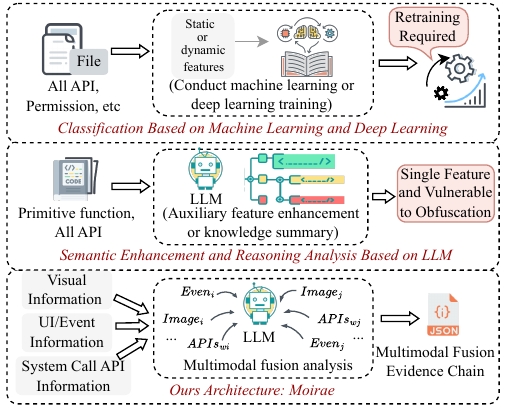}
\caption{Comparison of different scheme architectures.}
\label{fig:Scheme_comparison}
\end{figure}

\section{Related Work}
\label{bg}
As illustrated in Figure \ref{fig:Scheme_comparison}, static analysis based on superficial features such as API calls, permissions, and system call traces has been widely used to characterize the critical behaviors of Android applications \cite{cui2023api2vec, kim2018multimodal, qiu2022cyber, tam2015copperdroid, afonso2016going}. However, the continuous evolution of malware implementation techniques has triggered the problem of concept drift. To mitigate this issue, existing studies have explored package-level abstractions \cite{onwuzurike2019mamadroid}, semantic embeddings \cite{xu2020sdac, zhang2022slowing}, and model updating mechanisms that combine pseudo-labels with active learning \cite{xu2019droidevolver, chen2023continuous}. Simultaneously, dynamic analysis methods that collect system call traces, network traffic, and runtime API sequences within controlled sandboxes have also been widely introduced \cite{tam2015copperdroid, afonso2016going, wong2016intellidroid, tang2024dtdroid, feng2025hgdetector}. Although these methods enhance model resilience to a certain extent, they inherently still rely on fitting the feature distributions of historical data. When confronted with unknown threats exhibiting significant alterations in superficial features, such feature-fitting-based models often struggle to generalize and require continuous data collection for costly model retraining.

Leveraging LLMs to counter cybersecurity threats is emerging as a highly promising technical trajectory \cite{wang2024rethinking}. However, existing LLM application paradigms are frequently confined to code or single-dimensional features, making it challenging to acquire complete and mutually corroborating behavioral evidence. Under the augmentation paradigm, such as AppPoet \cite{zhao2025apppoet} and SRDC \cite{zhou2025srdc}, LLMs serve merely as semantic augmenters \cite{yan2025prompt}; the system ultimately remains constrained by the bottleneck of continuous retraining required for downstream learning-based models. Under the direct reasoning paradigm, such as AV-Agent \cite{zheng2025av}, LAMD \cite{qian2025lamd}, and MARD \cite{zeng2026mard}, approaches excessively rely on lengthy decompiled code. Their single-modal inputs lack cross-validation and fail to provide a comprehensive chain of evidence for complex, malicious behaviors.

\section{{\sysname} Architecture Overview}
\label{main}

\begin{figure*}[htbp]
\centering
\includegraphics[width=0.95\linewidth]{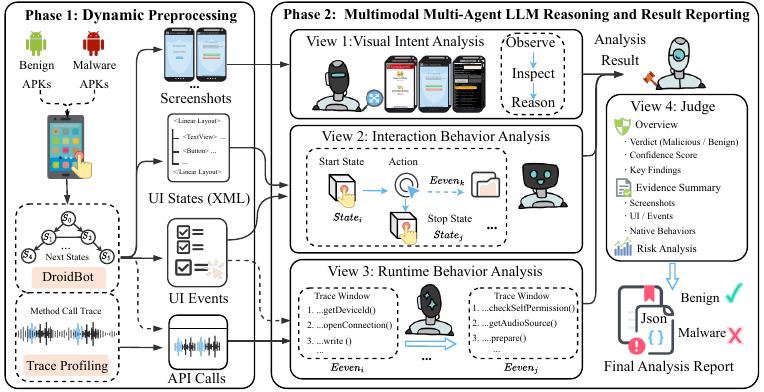}
\caption{Overview of the {\sysname} Architecture.}
\label{fig:Moirae}
\end{figure*}

Despite the constant changes in code and API combinations, the malicious targets behind them remain consistent. We propose {\sysname}, a multimodal agent collaborative framework for dynamic Android malware detection. The core idea of {\sysname} is to capture the runtime multimodal feature set $\mathcal{F} = \{\mathcal{V}, \mathcal{S},\mathcal{E}, \mathcal{T}\}$ of an application $\mathcal{A}$ across multiple semantic levels through dynamic execution. It establishes a precise causal mapping between superficial user interactions and underlying system API calls, and utilizes a set of LLM agents $Agents = \{A_{vis}, A_{ue}, A_{api}, A_{judge}\}$ for reasoning. As illustrated in Figure \ref{fig:Moirae}, the overall architecture comprises the following three key phases:

\textbf{Dynamic Multimodal Data Collection.} The system automatically drives the application under test $\mathcal{A}$ within a realistic Android emulator environment. During execution, the system synchronously records the device screenshot $\mathcal{V}$, captures the XML state tree sequence $\mathcal{S}$ containing detailed UI node information, and extracts the underlying ART (Android Runtime) method invocation trace $\mathcal{T}$ via an Android runtime profiling tool.

\textbf{Cross-Modal Causal Evidence Fusion.} The system aligns the timestamps of the semantic event set $\mathcal{E}$ generated during dynamic exploration with the underlying binary execution trace $\mathcal{T}$. By defining a precise time window $\Delta t$, {\sysname} accurately binds underlying API call sequences to the superficial UI interaction events $e_i \in \mathcal{E}$ that triggered them, thereby constructing a noise-eliminated cross-modal causal evidence $\mathcal{G}_{evidence}$.

\textbf{Multi-Agent Collaborative Reasoning and Decision-Making.} {\sysname} adopts a strategy of parallel independent reasoning followed by aggregated adjudication. We designed three parallel domain-expert agents that respectively retrieve specific modal evidence from the fusion report via a customized toolchain and conduct reasoning. Ultimately, an independent judge agent $A_{judge}$ aggregates the observation vectors from all experts to output the final threat classification and a highly interpretable, structured chain of evidence.

\section{Methodology Implementation}
\label{method}

\subsection{Dynamic Driving and Multimodal Feature Capture}
To trigger potential malicious payloads within the application, {\sysname} implements a highly automated dynamic sandbox environment driver. The system integrates the DroidBot \cite{li2017droidbot} engine and employs a Depth-First Search greedy algorithm-based strategy for autonomous Graphical User Interface (GUI) traversal, generating an interactive exploration action sequence $Action = \langle a_1, a_2, \dots, a_n \rangle$. During exploration, the driver continuously captures high-framerate screen recordings $\mathcal{V}$ via \texttt{screenrecord} and periodically exports the XML semantic state $\mathcal{S}_i$ of the UI. Simultaneously, the system utilizes the \texttt{am profile} command to launch the ART Profiler at the application process level, recording detailed method invocation Trace files $\mathcal{T}$ with low overhead. To ensure the reliability of the analysis results, the system executes deep package uninstallation, process termination, and device state snapshot rollback and cleanup operations after each analysis session.

\subsection{Cross-Modal Causal Evidence Fusion}
\label{Intent permission alignment}

Traditional dynamic analysis often generates massive volumes of underlying logs, which can easily trigger context overflow and hallucinations in LLMs. Therefore, we designed this module to achieve causal alignment and dimensionality reduction between underlying traces and high-level semantics. For every causal event $e_k \in \mathcal{E}$ captured by DroidBot, the fusion module extracts its timestamp $t_k$ and frames a time window within the underlying ART Trace $\mathcal{T}$ extending $\alpha$ seconds forward and $\beta$ seconds backward, denoted as $W_k = [t_k - \alpha, t_k + \beta]$. When parsing binary Trace records, the system precisely filters method invocation sequences by introducing a predefined whitelist set $\Omega$. This whitelist retains system APIs while filtering out customized APIs. For critical APIs satisfying the mapping function $f(api_j) \in \Omega$, the system not only records their execution order upon first invocation but also tallies their total trigger frequency $Freq(api_j, W_k)$ within the event time window $W_k$, ultimately outputting a high-density fusion report.

\subsection{Multi-Agent Collaborative Reasoning Based on LLMs}
\label{Context traceability}

To avoid performance bottlenecks in LLM when dealing with complex multimodal long contexts, {\sysname} constructed a multi-agent collaboration method based on the ReAct (Reasoning and Acting) paradigm.

\textbf{Superficial Visual Modality Insight.} The Vision Agent ($A_{vis}$) focuses on the global scanning and in-depth scrutiny of the application's GUI. Operating under the ReAct framework, this agent autonomously retrieves and traverses all screenshot evidence $\mathcal{V}$ captured during the dynamic execution phase. Through multimodal visual inference, behaviors such as endless pop-up ads, forced account password requests, and threatening and extorting interfaces were discovered. The unstructured pixel matrix was transformed into a structured visual risk feature representation $R_ {vis} $.

\textbf{Semantic Interaction Causal Reconstruction.} The UI/Event Agent ($A_{ue}$) conducts high-dimensional temporal analysis on the XML view tree structure and component interaction causal events. It concentrates on mining semantic misalignments between visible UI intentions and actual behaviors, silent malicious triggers requiring no user intervention, and anomalous privilege escalation trajectories, providing the system with a pure business-logic perspective feature representation $R_{ue}$ detached from underlying code.

\textbf{Deep Native Trace Forensics.} The API Agent ($A_{api}$) is responsible for executing micro-level dynamic forensics on concealed malicious payloads. This agent retrieves the native API invocation trace $T(W_k)$ strictly bound to each front-end event $e_k$. By executing context tracing across massive logs to identify malicious behaviors, it condenses the underlying physical behavior of the application into a definitive system-level abuse feature representation $R_{api}$.

\textbf{Cross-Modal Evidence Fusion and Adjudication.} Once all three parallel evidence agents complete their panoramic ReAct forensics, the Judge Agent ($A_{judge}$) fuses the evidence $\mathcal{M}_{judge}(R_{vis}, R_{ue}, R_{api})$ to analyze the asynchronous observation result vectors across the three major dimensions of visual perception, semantic interaction, and underlying traces. It makes the final adjudication for the APK, outputting an evaluation tuple $Decision = \langle y, C, S, E_{chain} \rangle$, which contains a Boolean malicious verdict $y \in \{0, 1\}$, a specific threat family classification $C$, a confidence score $S \in [0, 100]$, and a highly logically interpretable structured cross-validation evidence chain $E_{chain}$.

\section{Experimental Evaluation}
\label{experiment}

\begin{table*}[htbp]
\centering
\fontsize{9pt}{10.8pt}\selectfont
\begin{tabular}{@{} l cccc cccc cccc @{}}
\toprule
\multirow{2}{*}{\textbf{Approach}} 
& \multicolumn{4}{c}{\textbf{AndroZoo 2017--2021}} 
& \multicolumn{4}{c}{\textbf{CICMalDroid 2020}} 
& \multicolumn{4}{c}{\textbf{CIC-AndMal2017}} \\
\cmidrule(lr){2-5} \cmidrule(lr){6-9} \cmidrule(l){10-13}
& \textbf{ACC} & \textbf{Pre} & \textbf{Rec} & \textbf{F1} 
& \textbf{ACC} & \textbf{Pre} & \textbf{Rec} & \textbf{F1} 
& \textbf{ACC} & \textbf{Pre} & \textbf{Rec} & \textbf{F1} \\
\midrule

MalScan (KNN-1) & 72.60 & 71.55 & 73.26 & 72.40 & 60.90 & 63.88 & 86.15 & 73.36 & 45.57 & 39.33 & 74.27 & 51.43 \\
MalScan (KNN-3) & 75.91 & 74.27 & 77.83 & 76.01 & 58.01 & 62.31 & 83.08 & 71.21 & 44.82 & 39.31 & 77.67 & 52.20 \\
MalScan (RF)    & 61.73 & 65.44 & 46.52 & 54.38 & 25.00 & 37.74 & 30.77 & 33.90 & 39.17 & 33.97 & 60.19 & 43.43 \\
DroidEvolver    & 82.48 & \underline{85.98} & 77.75 & \underline{81.66} & 65.06 & 66.29 & 89.74 & 76.25 & 48.96 & 41.90 & \underline{81.55} & 55.35 \\
MaMaDroid       & 78.43 & \textbf{87.38} & 67.27 & 75.72 & 81.86 & \underline{86.48} & 84.92 & 85.58 & \underline{76.46} & \underline{70.56} & 68.98 & \underline{69.76} \\
CL-Malware      & \underline{84.88} & 84.29 & \textbf{85.86} & \textbf{85.07} & \underline{84.84} & 81.20 & \textbf{98.45} & \underline{88.99} & 67.23 & 55.52 & 78.16 & 64.92 \\
\midrule
\textbf{Moirae (Ours)} & \textbf{89.22} & 76.47 & \underline{83.07} & 79.58 & \textbf{90.06} & \textbf{91.00} & \underline{93.33} & \textbf{92.15} & \textbf{85.96} & \textbf{82.89} & \textbf{86.59} & \textbf{84.70} \\
\bottomrule
\end{tabular}
\caption{Cross-dataset generalization performance comparison (\%). We trained all baseline models on the AndroZoo (2011--2016) dataset. We tested them on unseen distribution datasets from different time periods (The Judge Agent uses the \texttt{GLM-5.1} model, while other agents use the \texttt{MiniMax-M3} model). The highest value is \textbf{bold}. The second-highest value is \underline{underlined}.}
\label{tab:cross_dataset_evaluation_horizontal}
\end{table*}

\subsection{Experimental Setup}
All experiments were implemented using Python 3.11 and conducted in an environment equipped with an Intel Core i7-12700 CPU and 32 GB of RAM. All LLMs utilized (including the MiniMax, Gemini, GLM, and Qwen series) were the official pristine versions \footnote{The GLM-5.1 models were invoked via the API interface provided by SiliconFlow (\url{https://siliconflow.cn/}), while the Gemini models were accessed via  OpenRouter (\url{https://openrouter.ai/}).}; parameters such as temperature were set to their defaults without any domain-specific fine-tuning. By default, View 1, View 2, and View 3 modules use \texttt{MiniMax-M3}. Meanwhile, the \texttt{Qwen3.6-27b} model is deployed locally on a single NVIDIA A100 (80GB VRAM) GPU server. Additionally, we conducted dynamic analysis based on the Android Emulator, dynamically executing each APK for 2 minutes. Code will be released upon acceptance.

\subsubsection{Datasets}
We selected three representative datasets: AndroZoo \cite{Allix:2016:ACM:2901739.2903508}, CICMalDroid 2020 \cite{mahdavifar2020dynamic}, and CIC-AndMal2017 \cite{lashkari2018toward}.
For AndroZoo, we randomly sampled applications spanning from 2011 to 2021. Combined with VirusTotal detection results (applications flagged by $\ge 10$ engines were labeled as malicious, while those with $0$ flags were labeled as benign), we constructed a sub-dataset comprising 16,216 applications (8,193 malicious and 8,023 benign). Conversely, CICMalDroid 2020 and CIC-AndMal2017 were strictly reserved as unknown test sets to challenge the cross-domain spatial robustness of the models.

\subsubsection{Baseline Models}
We compared {\sysname} with two categories of state-of-the-art (SOTA) models. \textbf{Static Graph/Behavior Models:} MaMaDroid \cite{onwuzurike2019mamadroid} models behavior transition probabilities via Markov chains. Malscan \cite{wu2019malscan} introduces centrality analysis from social networks into function call graphs. \textbf{Adaptive and Drift-Resistant Models:} DroidEvolver \cite{xu2019droidevolver} utilizes pseudo-labels for online model updates. CL-Malware \cite{chen2023continuous} combines active learning and continuous learning mechanisms to continuously accumulate knowledge.

\subsection{Research Questions (RQs)}

\begin{itemize}
\item \textbf{RQ1 (Effectiveness Analysis)}: How does {\sysname} perform overall? How does it compare to traditional baseline models relying on large-scale data training?


\item \textbf{RQ2 (Temporal Robustness):} How robust is {\sysname} to temporal concept drift without target-domain fine-tuning, compared with existing baselines?

\item \textbf{RQ3 (Ablation Study)}: How do different LLMs affect the final judgment accuracy? What role do the multi-dimensional evidence extraction stages play?


\item \textbf{RQ4 (Token Efficiency):} What token overhead does each analysis stage incur, and how effectively does hierarchical evidence extraction compress raw runtime traces?

\end{itemize}

\subsubsection{Evaluation Metrics}Based on the True Positives (TP), True Negatives (TN), False Positives (FP), and False Negatives (FN) from the confusion matrix, we calculate Accuracy (ACC), Precision (Pre), Recall (Rec), and the F1 score (F1). The specific calculations are presented in Table \ref{tab:evaluation_metrics}.

\begin{table}[htbp]
\centering
\resizebox{\linewidth}{!}{%
\scriptsize
\begin{tabular}{ll}
\toprule
Metric & Formula   \\
\midrule
Accuracy & $(\text{TP} + \text{TN})/(\text{TP} + \text{TN} + \text{FP} + \text{FN})$ \\
Precision  & $ (\text{TP})/(\text{TP} + \text{FP})$ \\
Recall  & $ (\text{TP})/(\text{TP} + \text{FN})$ \\
F1-score  & $ 2 \cdot (\text{Precision} \cdot \text{Recall})/(\text{Precision} + \text{Recall})$ \\
\bottomrule
\end{tabular}
}
\caption{Evaluation Metrics.}
\label{tab:evaluation_metrics}
\end{table}

\subsection{RQ1 -Effectiveness Analysis}
As shown in Table \ref{tab:cross_dataset_evaluation_horizontal}, we compare {\sysname} with conventional baseline models under a cross-dataset evaluation setting. The results show that {\sysname} maintains strong and consistent detection performance on previously unseen datasets, indicating improved robustness to distribution shifts and cross-dataset generalization.

Traditional machine learning and deep learning baselines rely primarily on shallow statistical features and static signatures learned from the AndroZoo training set collected between 2011 and 2016. Their performance degrades substantially when evaluated on datasets with markedly different distributions or large temporal gaps. For example, MalScan (RF) achieves only 25.00\% accuracy on CICMalDroid 2020, which contains more recent malware variants. On CIC-AndMal2017, all KNN- and RF-based baselines obtain accuracies below 50\%. Even the strongest baseline, CL-Malware, reaches only 67.23\% accuracy and a 64.92\% F1 score on this dataset. These results highlight the limited temporal and cross-dataset generalization of conventional supervised detectors in the face of malware technological evolution.

In contrast, {\sysname} achieved distinct advantages across various unseen datasets. It accomplished this without any target-domain data fine-tuning. On an unseen dataset, the average ACC is 88.01\%, and F1 is 88.43\%, which is a relative improvement of 9.13\% and 11.41\% compared to the best baseline. This strongly proves that the multi-agent architecture based on LLMs is no longer limited to memorizing underlying binary feature distributions. Instead, it possesses deep semantic understanding and causal reasoning capabilities, and captures the core motives of malware deception and theft.

\begin{mybox}
\textbf{Conclusion for RQ1:} {\sysname} effectively mitigates cross-dataset generalization degradation in Android malware detection. It achieves an average ACC of 88.01\% and F1 of 88.43\% on unseen data distributions without fine-tuning, surpassing the best baseline by 9.13\% in ACC. This demonstrates its robust generalization to unseen malware variants.
\end{mybox}

\subsection{RQ2: Temporal Robustness}

We evaluate {\sysname} under temporal concept drift using chronologically organized AndroZoo samples from 2017 to 2021. Baseline models are trained on historical 2011–2016 samples, while {\sysname} performs zero-shot inference.

\begin{figure}[h]
\centering
\includegraphics[width=0.95\linewidth]{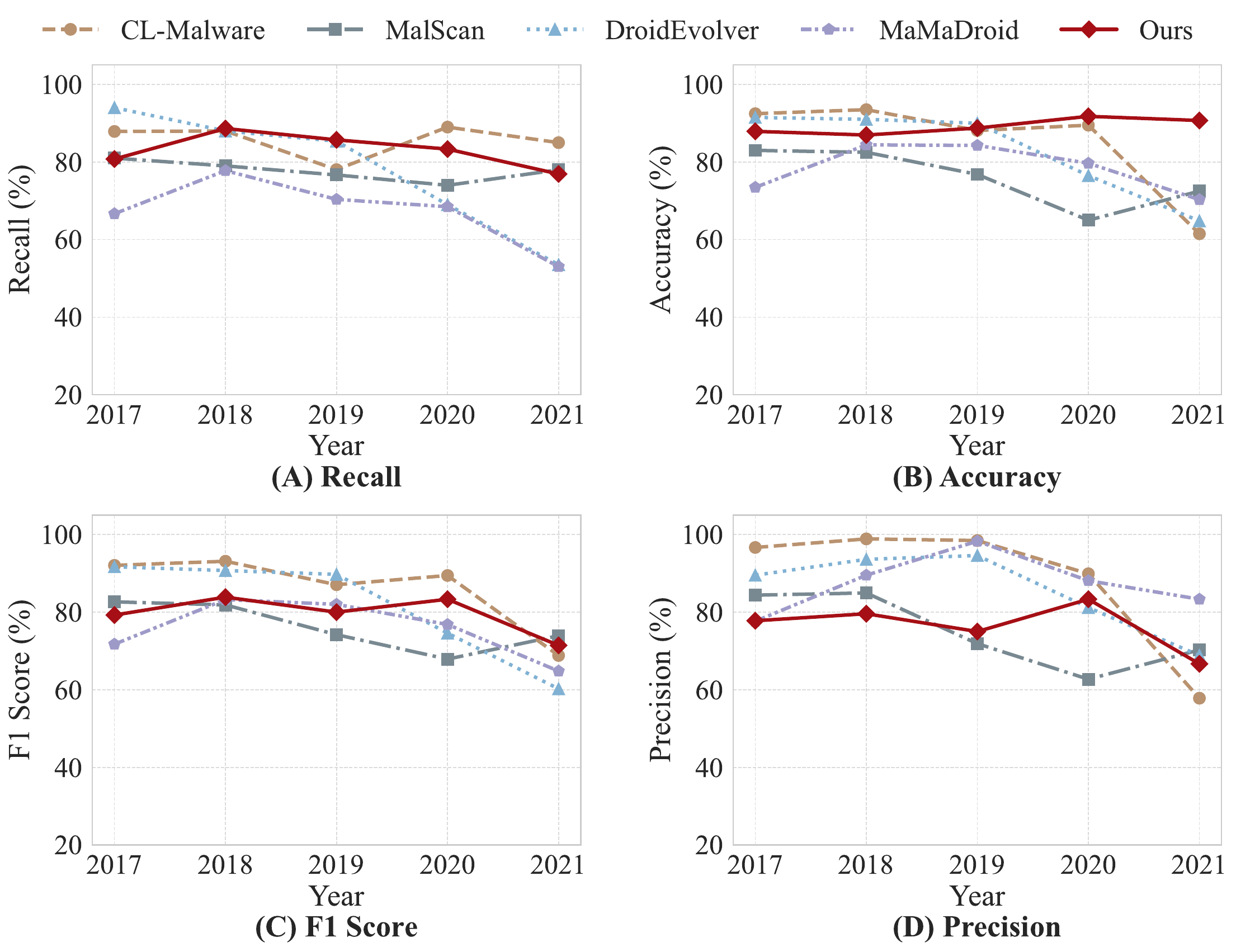}
\caption{Time-aware evaluation based on AndroZoo.}
\label{fig:AndroZoo_year}
\end{figure}

As shown in Figure~\ref{fig:AndroZoo_year}, the baselines generally degrade as the temporal gap increases. For example, the accuracy of CL-Malware decreases from 92.46\% in 2017 to 61.50\% in 2021, while the F1 score of DroidEvolver drops from 91.71\% to 60.23\%. This degradation occurs because supervised detectors learn feature-label associations from historical samples. These associations become outdated as Android ecosystems, APIs, libraries, and malware implementations evolve. In comparison, the accuracy of {\sysname} remains within approximately 87\%--92\% throughout the five years and stays around 90.70\% in 2021. 



Nevertheless, {\sysname} is still affected by the inherent limitations of dynamic analysis. Its recall decreases to approximately 77\% in 2021 because conditionally triggered or environment-dependent malicious behaviors may not be activated during dynamic exploration. Moreover, partially observed data may lack sufficient context, causing benign but sensitive operations to be interpreted as malicious and thereby reducing precision. 

\begin{mybox}
\textbf{Conclusion for RQ2:}
{\sysname} substantially mitigates temporal degradation without fine-tuning. It maintains approximately 90\% accuracy on the 2021 samples, outperforming baselines by approximately 18--29 percentage points. Its remaining performance loss primarily arises from the inherent difficulty of triggering and observing complete malicious behaviors during dynamic analysis.
\end{mybox}

\subsection{RQ3 - Ablation Study}

We designed two groups of ablation studies to investigate the impact of different design choices in {\sysname}. The first study examines how the choice of LLM for the Judge Agent affects detection performance, while the second evaluates the contribution of evidence modules from View 1 to View 3.

\begin{table}[h]
\centering
\resizebox{\linewidth}{!}{
\begin{tabular}{@{}llcccc@{}}
\toprule
\makecell{\textbf{Evidence} \\ \textbf{Agents}} 
& \makecell{\textbf{Judge} \\ \textbf{Agent}} 
& \textbf{ACC} & \textbf{Pre} & \textbf{Rec} & \textbf{F1} \\
\midrule

\multirow{4}{*}{MiniMax-M3}
& MiniMax-M3       & 88.93 & 91.76 & 90.17 & 90.96 \\
& Gemini-3.1-Pro   & 85.26 & \textbf{92.09} & 83.59 & 87.63 \\
& GLM-5.1          & \textbf{90.06} & 91.00 & 93.33 & \textbf{92.15} \\
& GLM-4.7-Flash    & 87.18 & 91.44 & 87.69 & 89.53 \\
\midrule

\multirow{4}{*}{Qwen3.6-27b}
& Qwen3.6-27b      & 83.65 & 80.51 & \textbf{97.44} & 88.17 \\
& GLM-5.1          & 85.25 & 82.53 & 96.92 & 89.15 \\
& GLM-4.7-Flash & 78.14 & 75.20 & 96.91 & 84.68 \\
& MiniMax-M3       & 84.89 & 84.51 & 92.78 & 88.45 \\
\bottomrule
\end{tabular}
}
\caption{Multi-agent model ablation experimental results on the CICMalDroid 2020 dataset (\%). The View 1 to View 3 modules (evidence extraction) utilize \texttt{MiniMax-M3} or the locally deployed \texttt{Qwen3.6-27b}.}
\label{tab:judge_agent_ablation}
\end{table}

As shown in Table \ref{tab:judge_agent_ablation}, different LLMs exhibit distinct decision behaviors when serving as the Judge Agent. In our framework, \texttt{MiniMax-M3} is used for evidence extraction, while combining it with \texttt{GLM-5.1} as the reasoning layer achieves the best overall performance, reaching 90.06\% accuracy and 92.15\% F1 score.



\begin{table}[h]
\centering
\fontsize{9pt}{10.8pt}\selectfont
\begin{tabular}{@{}ccccccc@{}}
\toprule
\multicolumn{3}{c}{\textbf{Evidence Modules}} & \multicolumn{4}{c}{\textbf{Performance (\%)}} \\
\cmidrule(r){1-3} \cmidrule(l){4-7}
\textbf{View 1} & \textbf{View 2} & \textbf{View 3} & \textbf{ACC} & \textbf{Pre} & \textbf{Rec} & \textbf{F1} \\
\midrule

\checkmark & \checkmark & \checkmark & \textbf{88.93} & 91.76 & \textbf{90.17} & \textbf{90.96} \\
\midrule

& \checkmark & \checkmark & 85.95 & 90.50 & 86.63 & 88.52 \\
\checkmark &  & \checkmark & 78.82 & 93.38 & 70.95 & 80.63 \\
\checkmark & \checkmark &  & 86.05 & 91.76 & 85.25 & 88.39 \\
\midrule

\checkmark &  &  & 78.32 & 95.04 & 69.07 & 80.00 \\
& \checkmark &  & 86.32 & 89.84 & 87.96 & 88.89 \\
&  & \checkmark & 70.16 & \textbf{95.37} & 54.50 & 69.36 \\
\bottomrule
\end{tabular}
\caption{Evidence module ablation experiments on the CICMalDroid 2020 dataset. The \checkmark indicates that the experiment retained the evidence input of that module.}
\label{tab:module_ablation}
\end{table}


We also observe a clear precision-recall tradeoff among different Judge Agents. \texttt{Gemini-3.1-Pro} adopts a more conservative decision strategy, achieving the highest precision (92.09\%) while sacrificing some recall. In contrast, the locally deployed open-source model \texttt{Qwen3.6-27b} exhibits a more sensitive detection tendency, achieving the highest recall (97.44\%) at the cost of lower precision. These results demonstrate the flexibility of {\sysname} in balancing precision and recall through different Judge Agents for diverse security scenarios.

Table \ref{tab:module_ablation} further analyzes the contribution of each component in the cross-modal evidence chain. The results show that integrating all three views achieves the strongest overall detection performance. Among them, View 2 plays an important role. In the "remove single module" setting, removing View 2 leads to the largest performance degradation, reducing accuracy by 10.11 percentage points to 78.82\%. Meanwhile, in the "retain single module" setting, View 2 alone maintains an accuracy of 86.32\%, indicating that UI interaction states provide highly discriminative behavioral evidence for malware detection.

In contrast, using only View 3 achieves the highest precision (95.37\%) but suffers from a substantially reduced recall (54.50\%). This suggests that low-level API evidence alone is insufficient for comprehensive detection, as sophisticated malware may conceal malicious behaviors through techniques such as reflection and dynamic loading. Incorporating high-level semantic information from View 1 and View 2 helps establish the connection between user-visible behaviors and underlying sensitive operations.

\begin{mybox}

\textbf{Conclusion for RQ3:} {\sysname} benefits from heterogeneous LLM agents with complementary decision preferences, where different Judge Agents achieve flexible Pre-Rec trade-offs (e.g., \texttt{Gemini-3.1-Pro} achieves 92.09\% Pre, while \texttt{Qwen3.6-27b} reaches 97.44\% Rec). Multi-dimensional evidence fusion is essential for robust detection, with View 2 playing a central role, and full views achieving the best overall performance.
\end{mybox}

\subsection{RQ4 - Token Efficiency}
To evaluate the deployment token efficiency of {\sysname}, we analyze its token consumption during malware analysis across different datasets and malware families.

\begin{figure}[h]
\centering
\includegraphics[width=0.95\linewidth]{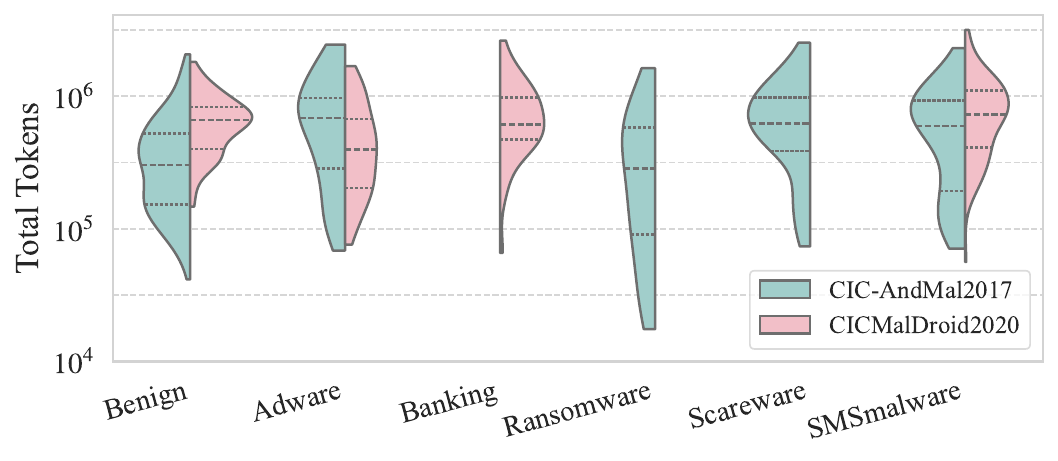}
\caption{Distribution of System Total Tokens by Family and Dataset.}
\label{fig:Token_distribution}
\end{figure}

\begin{figure}[h]
\centering
\includegraphics[width=0.95\linewidth]{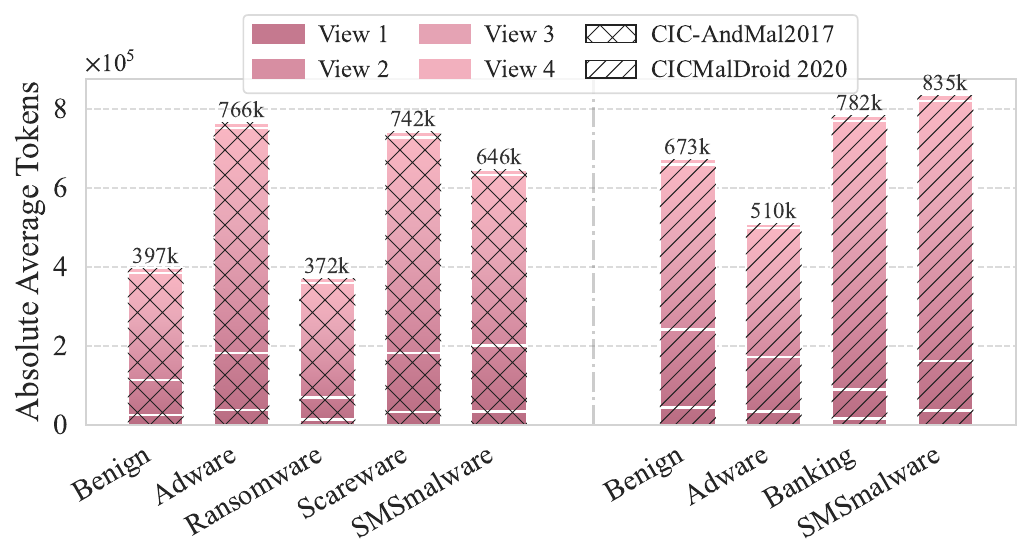}
\caption{Absolute Token Cost and Agent Contribution Across Datasets.}
\label{fig:Cost_contribution}
\end{figure}

As shown in Figure \ref{fig:Token_distribution}, the overall token consumption mainly falls within the range of $10^5$ to $10^6$ tokens. Different malware families exhibit distinct consumption patterns due to their behavioral characteristics. Specifically, Figure \ref{fig:Cost_contribution} presents the average token consumption across categories. On the CIC-AndMal2017 dataset, Adware (766k), Scareware (742k), and SMSmalware (646k) consume substantially more tokens than Benign (397k) and Ransomware (372k) samples. Similarly, on CICMalDroid 2020, SMSmalware incurs the highest average consumption (835k), while Benign samples remain relatively stable at approximately 673k tokens. These differences are consistent with the behavioral complexity of different malware categories. For example, Adware and SMS-based malware frequently trigger UI interactions and extensive API call sequences, resulting in longer behavioral traces. In contrast, Ransomware typically performs rapid file-encryption operations shortly after execution, producing shorter behavioral trajectories and requiring fewer evidence tokens.

\begin{figure}[h]
\centering
\includegraphics[width=0.95\linewidth]{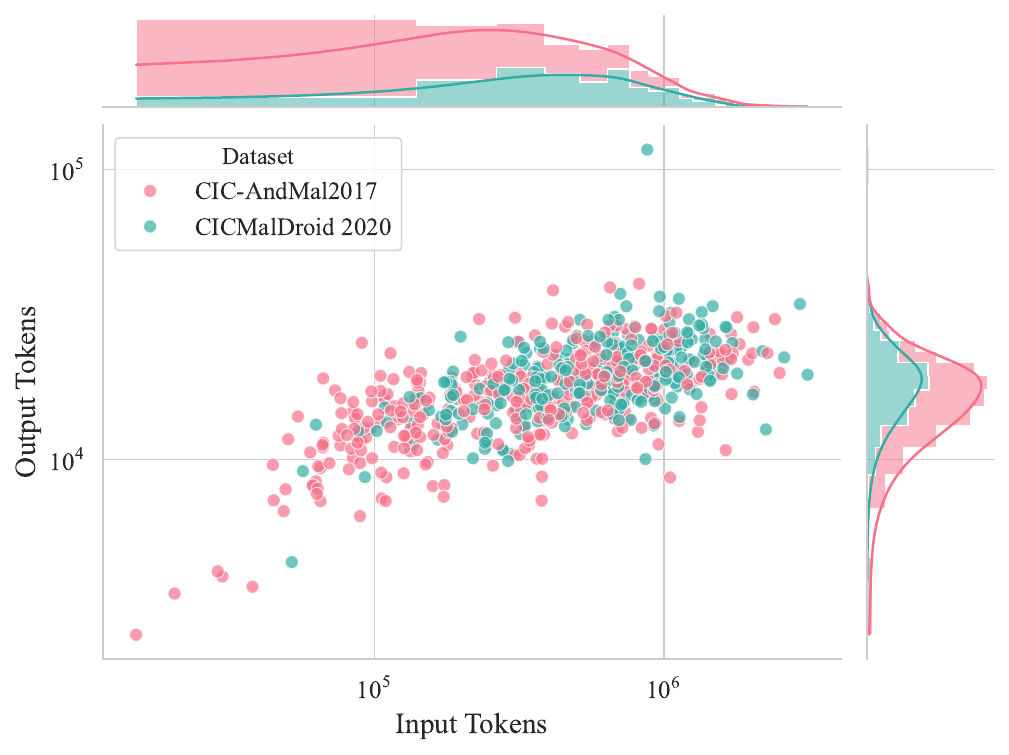}
\caption{Consumption of Input and Output Tokens.}
\label{fig:in_output_token}
\end{figure}

Figure \ref{fig:Cost_contribution} further illustrates the token contribution of different View agents. View 3 contributes the majority of the total token overhead, mainly because low-level API sequences and system-call traces naturally contain highly repetitive and fine-grained information. In comparison, the Judge Agent consumes only a small fraction of the total tokens after receiving the compressed evidence representations. This demonstrates the effectiveness of the evidence extraction method in reducing redundant low-level information. Without such intermediate abstraction, directly providing raw API sequences to the Judge Agent would likely exceed the context window and introduce substantial inference latency and computational cost.

Figure \ref{fig:in_output_token} further highlights the effectiveness of {\sysname}. In the logarithmic scatter plot, input tokens are broadly distributed in the range of $10^5$ to $10^6$, whereas output tokens from the evidence extraction agents are concentrated around $10^4$. This indicates that Views 1--3 effectively transform lengthy behavioral traces into compact semantic representations. The resulting compression ratio ranges from approximately 10$\times$ to 100$\times$, enabling the Judge Agent to perform efficient reasoning within a constrained context window.

\begin{mybox}

\textbf{Conclusion for RQ4:} {\sysname} achieves efficient token utilization by adapting consumption to malware complexity and compressing raw behavioral logs by 10$\times$--100$\times$. Despite higher costs for interaction-intensive malware families, it enables scalable LLM-based analysis under constrained context windows.
\end{mybox}

\section{Conclusion}
\label{conclusion}


This paper presents {\sysname}, an LLM-powered multimodal agent framework for Android malware detection under concept drift. Instead of relying on implementation-specific features or single-dimensional evidence, {\sysname} dynamically integrates visual inducements, UI interaction transitions, and system-level API operations to reconstruct cross-level behavioral chains and infer high-level malicious intent. Extensive experiments on temporally and distributionally unseen datasets show that {\sysname} achieves zero-shot generalization without target-domain fine-tuning, reaching an accuracy of 90.06\% and outperforming SOTA baselines.

\bibliographystyle{unsrt}
\bibliography{myref}


\end{document}